\documentclass{article}
\usepackage{nips2003e,times,epsfig}

\title{Ambiguous model learning made unambiguous with 1/f priors}

\author{G. S. Atwal \\
Department of Physics\\
Princeton University\\
Princeton, NJ 08544 \\
\texttt{gatwal@princeton.edu} \\
\And
William Bialek\\
Department of Physics\\
Princeton University\\
Princeton, NJ 08544\\
\texttt{wbialek@princeton.edu}
}

\begin{document}

\newcommand{\al}{\alpha}
\newcommand{\alb}{\overline{\alpha}}
\newcommand{\ml}{\overline{\Pi}_0(\qt)}
\newcommand{\nid}{\noindent}
\newcommand{\ds}{\displaystyle}
\newcommand{\ha}{\widehat}
\newcommand{\hb}{\hbar}
\newcommand{\si}{\sigma}
\newcommand{\la}{\lambda}
\newcommand{\ep}{\epsilon}
\newcommand{\cd}{\cdot}
\newcommand{\de}{\delta}
\newcommand{\na}{\nabla}
\newcommand{\om}{\omega}
\newcommand{\Om}{\Omega}
\newcommand{\da}{\dagger}
\newcommand{\lan}{\langle}
\newcommand{\ran}{\rangle}
\newcommand{\pder}[2]{\frac{\partial{#1}}{\partial {#2}}}
\newcommand{\cder}[1]{\frac{D {#1}}{Dt}}
\newcommand{\pdder}[2]{\frac{{\partial}^2{#1}}{\partial {#2}^2}}
\newcommand{\vder}[2]{\frac{\delta {#1}}{\delta {#2}}}
\newcommand{\vdder}[2]{\frac{\delta^2 {#1}}{\delta {#2}^2}}
\newcommand{\bearraynn}{\begin{eqnarray}}
\newcommand{\eearraynn}{\end{eqnarray}}
\newcommand{\benn}{\begin{equation}}
\newcommand{\eenn}{\end{equation}}
\newcommand{\eq}[1]{{Eq.~(\ref{#1})}}
\newcommand{\eqs}[2]{{Eqs.~(\ref{#1}--\ref{#2})}}
\newcommand{\lab}{\label}
\newcommand{\frd}[2]{\ds \frac{{#1}}{{#2}}}
\newcommand{\De}{\Delta}
\newcommand{\pxa}{P[x|\al(t)]}
\newcommand{\pxab}{P[x|\alb(t)]}
\maketitle

\begin{abstract}
What happens to the optimal interpretation of noisy data when there exists more than one equally plausible interpretation of the data? In a Bayesian model-learning framework the answer depends on the prior expectations of the dynamics of the model parameter that is to be inferred from the data. Local time constraints on the priors are insufficient to pick one interpretation over another. On the other hand, nonlocal time constraints, induced by a $1/f$ noise spectrum of the priors, is shown to permit learning of a specific model parameter even when there are infinitely many equally plausible interpretations of the data. This transition is inferred by a remarkable mapping of the model estimation problem to a dissipative physical system, allowing the use of powerful statistical mechanical methods to uncover the transition from indeterminate to determinate model learning.
\end{abstract}

\section{Introduction}
The estimation of a model underlying the production of noisy data becomes highly nontrivial when there exists more than one equally plausible model that could be responsible for the output data. The viewing of ambiguous figures, such as the Necker cube [1], is a classical problem of this type in the field of visual psychology. Pitch perception when hearing a number of different harmonics is another example of ambiguous perception [2].

Previous studies [3] have reduced the problem of optimal interpretation of an ambiguous stimulus to the problem of estimating a single variable which may vary in time $\al(t)$, given a time sequence of noisy data. Enforcing a prior belief that the local dynamics $\al(t)$ should not vary too rapidly embodies the observer's knowledge that rapid variations in $\al(t)$ are unlikely in the natural world or in a given experiment. Such a prior prevents overfitting the model estimate to the data as it arrives. The statistically optimal interpretation of the data was then found to consist of $\al(t)$ hopping randomly from one possible interpretation to another. The rate of random switching between interpretations was found to be controlled not by the noise level (e.g. in the neural hardware), as previously thought, but rather by the observer's prior hypotheses. This hopping persists indefinitely despite the fact that the probability distribution of the incoming data remains the same. In such cases it is impossible to learn a specific model parameter.

In this paper we introduce another prior over the dynamics of $\al(t)$. We assume that fluctuations in $\alpha (t)$ have a $1/f$ spectrum, as observed ubiquitously in nature. Such a prior is shown to induce nonlocal time constraints on the trajectories of $\al(t)$ and, unlike the local constraints, can result in specific model learning in the case of ambiguous models. The fact that $1/f$ priors can induce unambiguous model learning is the central result of this work.

The analyses of the long-time dynamics with nonlocal priors is permitted by a surprising and remarkable mapping to a dissipative quantum system. This mapping not only guides our intuition of the optimal trajectories of $\al(t)$ but also permits the usage of powerful statistical mechanical techniques. In particular, the renormalization group (RG) can be employed to uncover the conditions in which there is a transition from non-specific model learning to specific model learning.

\section{Formalism}

Suppose that we are given a series of $N$ measurements $\{ x_t \}$ at discrete times $t$. Then Bayes rule gives us the conditional probability of $\{ \al_t \}$ giving rise to those data
\benn
P[\{ \al_t \} | \{x_t\}]=\frac{P[\{x_t\}|\{ \al_t \} ] \, P[\{ \al_t \}]}{P[\{x_t\}]}, \lab{bayes}
\eenn
\nid where the probability of making the observations $\{x_i\}$ is given by summing up all the possible models that may give rise to them,
\benn
P(\{x_t\})=\int d\al \, P[\{x_t\}| \{ \al_t \} ] P[ \{ \al_t \} ].
\eenn
\nid We further assume conditional independence of signals,
\benn
P[\{x_t\}| \{ \al_t \} ]=P[x_1x_2...x_N| \{ \al_t \}] = \prod_{t=1}^N P[x_t|\al_t ]. \lab{px}
\eenn
A natural step is then to consider how close our estimate of the model $\al(t)$ lies to the true underlying model $\alb(t)$, which we take to be stationary $\alb(t)=\alb$. We can think of these probability distributions as Boltzmann
distributions in which some effective potential acts to hold $\alpha$
close to $\bar \alpha$; thus we envision an energy landscape in the
$\alpha$ space with a minimum at $\bar\alpha$. 

A more interesting, and generalized, question arises when we consider the global properties of the extended energy landscape. In particular there may be $M>1$ equally plausible interpretations consistent with the input data\footnote{Of course it may be the case that some interpretations may be more plausible than others, resulting in a non uniform probability distribution over possible models. In this paper we illustrate the case where all interpretations are equally likely, $P[\alb_m]=1/M$.} in which case there exist degenerate minima at $\alb_m$ ($m=1,2....M$),

\benn
P[x_t|\alb_1]=P[x_t|\alb_2]=...=P[x_t|\alb_M].
\eenn

Therefore we may write \eq{px} as
\benn
P[\{ x_t \} | \{ \al_t \} ] = \prod_{t=1}^N \left( \prod_{m=1}^M  P[x_t|\alb_m]^{1/M} \right) \exp \left[ \frac{1}{M} \sum_{m=1}^M \sum_{t=1}^N \ln \frac{P[x_t| \al_t]}{P[x_t|\alb_m]} \right].
\eenn
On average, the term in square brackets is related to the Kullback-Leibler divergences between distributions conditional on $\alpha (t)$ and distributions conditional on the true $\bar \alpha$. If the time variation of $\alpha$ is slow, we effectively collect many samples of $x$ before $\alpha$ changes, and it makes sense to replace
the sum over samples by its average:
\bearraynn
\lim_{N \to \infty} \sum_{m=1}^M \sum_{t=1}^{N} \ln \frac{P[x_t| \al_t]}{P[x_t|\alb_m]} &\approx& \frac{1}{\tau_0} \sum_{m=1}^M \int dt \int dx P[x(t)|\alb_m] \ln \frac{P[x(t)|\al(t)]}{P[x|\alb_m]}, \nonumber \\
&\equiv& -\frac{1}{\tau_0} \sum_{m=1}^M \int dt D_{KL} [\alb_m || \al(t)].
\eearraynn
\nid where $\tau_0$ is the average time between observations, and we take the continuum limit.

\subsection{Priors}

We need to have some prior hypotheses about how $\al(t)$ can vary in time, serving as our prior probability distribution $P[\al(t)]$. We introduce two different types of priors characterized by whether they constrain the local or nonlocal time dynamics,

\benn
P[\al(t)]=P_{\rm local}[\al(t)] P_{\rm nonlocal}[\al(t)].
\eenn

To summarize our prior expectation that the local dynamics of $\al(t)$ vary slowly, we assume that the time derivative of $\al(t)$ is chosen independently at each instant of time from a Gaussian distribution,
\benn
P_{\rm local}[\al(t)] \propto \exp \left[ -\frac{1}{4 D} \int dt \left( \pder{\al}{t} \right)^2 \right]. \lab{loc}
\eenn
\nid Note that this distribution corresponds to random walk with effective diffusion constant $D$.

Motivated by the ubiquitous occurrence of $1/f$ fluctuations in nature we chose to encapsulate the nonlocal dynamics by a Gaussian distribution with a $1/f$ power spectrum of noise, conveniently expressed in Fourier coordinates $\om$ as 
\benn
P_{\rm nonlocal}[\al(t)] \propto \exp \left[ - \frac{1}{2} \int \frac{d\om}{2\pi} \frac{|\al(\om)|^2}{S(\om)} \right], \lab{nonloc}
\eenn
\nid where the spectral noise function takes the form
\benn
S(\om)=\frac{1}{\eta |\om|}. \lab{1/f}
\eenn
\nid Note that the spectrum must be even in $\om$ since for any stationary process $S(\om)=S(-\om)$. The parameter $\eta$ determines the strength of {\it a priori} belief in nonlocal dynamics, or as we will see later, it can be equivalently viewed as a frictional constant determining the dissipation of the time trajectories of $\al(t)$. In the time-domain \eq{nonloc} becomes
\benn
P_{\rm nonlocal}[\al(t)] \propto \exp \left[ -\frac{\eta}{4 \pi} \int dt dt' \left( \frac{\al(t)-\al(t')}{t-t'} \right)^2 \right]. \lab{noi}
\eenn
Combining \eq{loc} and \eq{noi} we then obtain the total prior expectation of the probability distribution over the time-dependence of the model parameter $\al(t)$
\benn
P[\al(t)] \propto \exp \left[ -\frac{1}{4 D} \int dt \left( \pder{\al}{t} \right)^2 -\frac{\eta}{4 \pi} \int dt dt' \left( \frac{\al(t)-\al(t')}{t-t'} \right)^2 \right]. 
\eenn
Taken together, the local and non-local terms describe fluctuations in $\al$ which are $1/f$ up to a cutoff frequency, $\om_c \sim D \eta$. 
Returning to the Bayesian conditional probability \eq{bayes} we then obtain a path-integral expression 
\benn
P[\al(t)|\{ x_i \} ] \propto \exp (-S[\al(t)]),
\eenn
\nid where the action $S[\al(t)]$ is given by
\bearraynn
S[\al(t)]&=& \int dt \left[ \frac{1}{4D} \left( \pder{\al}{t} \right)^2 + \eta \int \frac{dt'}{4 \pi} \left( \frac{\al(t)-\al(t')}{t-t'} \right)^2  +  V_{\rm eff}[\al(t)] \right], \lab{path} \\
V_{\rm eff}[\al(t)] &=& \frac{1}{\tau_0 M} \sum_{m=1}^{M} D_{KL} [\alb_m || \al(t)].
\eearraynn
\nid This is equivalent to the imaginary time path-integral for a quantum mechanical particle [4] of mass $1/2D$ , with coordinates given by $\alpha(t)$, moving in an effective potential $V_{\rm eff}[\al(t)]$ and subject to (linear) frictional forces with a damping constant $\eta$. This mapping provides an extremely useful guide to our intuition for the probable trajectories of $\al(t)$. Just as in the analyses of particle dynamics in dissipative quantum mechanics [4] we anticipate that the time-course of $\alpha(t)$ may exhibit qualitatively different types of behavior depending on the strength of the non-local terms. In addition, the equivalence to a physical system permits exploitation of powerful techniques developed in the study of quantum mechanical systems with infinite degrees of freedom.

In the following we consider the cases of $m=1$ and $m=2$ and use the RG transformations to consider localization-delocalization transitions.

\subsection{M=1 : One true interpretation of data}

Now if $\al(t)$ differs from  $\alb$ by a small $\Delta \al(t)$ we can Taylor expand the Kullback-Leibler divergence to give a quadratic distance measure

\benn
D_{KL} (\alb || \al) =\frac{1}{2} F[\alb(t)] \Delta \al(t)^2 + O(\Delta \al^3),
\eenn

\nid where the metric is the Fisher information
\benn
F[\al(t)]=\int dx \frac{1}{\pxa} \left( \pder{\pxa}{\al(t)} \right)^2.
\eenn

Thus, close to the true parameter $\alb$ the potential energy term in \eq{path} is simply a harmonic oscillator with stiffness given by the Fisher information. Guided by the mapping to a dissipative quantum mechanical system we expect that if the initial distribution of $\alpha$ already happens to be closely centered around the correct value then the most likely trajectory will be simply to move closer to the minima of the potential energy at $\alb_1$. 

The important point to note is that had we chosen just the local constraints on our priors \eq{loc} then the trajectory of $\al(t)$ would persistently fluctuate around $\alb_1$, representing a trade-off between avoiding overfitting the data and inertia of our estimate. In the quantum mechanical picture this corresponds to the zero point fluctuations around the minima. Adding the dissipative term reduces the fluctuations around $\alb_1$ by an amount monotonically dependent on $\eta$, thus improving on the optimal estimate.

A RG treatment of the single-well problem, within the harmonic approximation, renormalizes the Fisher information such that the curvature of the potential well increases for all values of the $\eta$, and thus the fixed point of the dynamics is simply the convergence of $\al(t)$ to reduced fluctuations around the true parameter $\alb_1$. We explicitly carry out the RG calculation in the more interesting case where we have two global minima in the next section.

\subsection{M=2 : Two equally possible interpretations of the data}

In the case of two equally viable interpretations of the data, the potential energy term becomes that of a double-well potential with degenerate minima at $\alb_1$ and $\alb_2$ and energy barrier $h$ \benn
h=\frac{1}{2 \tau_0} \left( D_{KL}(\alb_1||(\alb_1+\alb_2)/2)+D_{KL}(\alb_2||\alb_1+\alb_2)/2) \right) 
\eenn

\begin{figure}[h]
\begin{center}
{\parbox{5cm}{\resizebox{6cm}{5cm}{\includegraphics{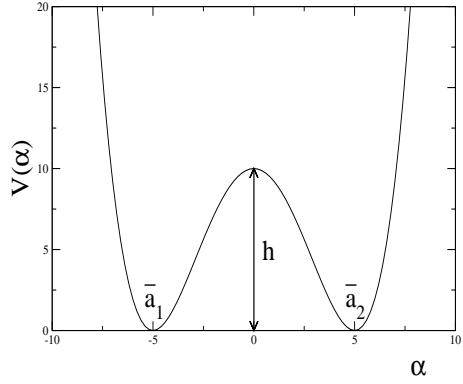}}}}
\end{center}
\caption{Potential energy landscape for $\al$ where there exist two equally valid interpretations. \eq{qu}} \lab{quart}
\end{figure}

\nid Without any dissipative dynamics, the optimal estimate of $\al(t)$ will switch between the two minima, representing instanton trajectories of a quantum particle tunnelling through the energy barrier backwards and forwards [3]. In contrast, it is well known that, at least in some regimes, the problem with dissipation has a phase transition to a truly localized state. Previous work has demonstrated such a dynamical phase transition in the strong-coupling limit (i.e. large barrier height limit) using semi-classical approximations for the dynamics [4,5,6], and in this section we will show that a perturbative RG treatment yields similar results in the opposite weak-coupling limit. 

For the sake of simplicity we employ the following simple quartic potential (see Fig.\ref{quart}), although the results will be independent of its exact form,
\benn
V(\al)=\frac{h}{\alb_1^4} \left( \al^2- \alb_1^2 \right)^2. \lab{qu}
\eenn
\nid The $\al$ coordinates have been shifted such that $\alb_1=-\alb_2$, and the height $h$ of the energy barrier located at $\al=0$ sets the overall energy scale. It is useful to write the effective action of \eq{path} in dimensionless parameters 
\benn
a=\frac{\al}{\alb_1}, \hspace{1cm}  b=\eta \alb_1^2, \hspace{1cm} c=\frac{h}{\Lambda},
\eenn
\nid where $\Lambda=D/ \alb_1^2$ is the energy/frequency scale \footnote{The constant of proportionality between energy and frequency is set to 1, akin to the common physics computation setting of $\hbar=1$.}
\bearraynn
S&=&\frac{1}{2} \int \frac{d \om}{2 \pi} \left( \frac{1}{2 \Lambda} \om^2+ b |\om| \right) |a(\om)|^2 + c \Lambda \int dt \, V'(a), \\
V'(a)&=&(a^2-1)^2. \label{dimpot}
\eearraynn

By power counting in the first integral the dissipative term, at low frequencies, dominates over the kinetic energy term. In the language of RG, the kinetic energy term is an irrelevant operator and can thus be ignored if we now focus our attention to frequencies below some cut-off $\lambda$. To determine the RG flow of the dimensionless coupling parameters the high-frequency components are integrated out from $\om=\lambda-d\lambda$ to $\om=\lambda$ to give a new effective action $\tilde{S}$ over the low frequency modes $\om< \lambda$. To accomplish this the function $\al(\om)$ is split
\benn
a(\om)=a_< (\om) \theta(|\om| < \lambda - d\lambda) + a_> (\om) \theta(\lambda-d \lambda < |\om| < \lambda), 
\eenn
and the new action is obtained by integrating over $a_>(\om)$,
\bearraynn
Z&=&\int {\cal D} a \exp [-S(a)], \nonumber \\
&=& \int \int {\cal D} a_< {\cal D} a_> \exp[-S(a_< + a_>)], \nonumber \\
&=& \int {\cal D} a_< \exp[-\tilde S(a_<)].
\eearraynn
Therefore,
\benn
\tilde{S}(a_<)=\frac{b}{2} \int_0^{\lambda- d \lambda} \frac{d \om}{2 \pi} |\om| |a_< (\om)|^2 + \ln \left< \exp \left[ c \Lambda \int dt V'(a_< + \al_>) \right] \right>_{a_>}, \lab{sa}
\eenn
\nid where the averaging is defined by
\benn
\lan A \ran_{a_>} \propto \int {\cal D} a_> \exp \left\{ -\frac{b}{2}  \int_{\Lambda-d \Lambda}^{\Lambda} \frac{d\om}{2 \pi} |\om| |a_>(\om)|^2 \right\} A.
\eenn
In the weak-coupling limit, we may expand the exponential term in \eq{sa} before performing the averaging,
\benn
\left< \exp[c \Lambda \int dt V'(a_<+a_>)] \right>_{a_>}= \left< 1 + c \Lambda \int dt V'(a_<+a_>)+...\right>_{a_>}. 
\eenn
Terminating the expansion to first order in the potential represents a one-loop calculation in field theories. 

Making use of
\benn \left< a_>^2(t) \right>_{a_>} = \int_{\lambda-d\lambda}^{\lambda} \frac{d \om}{\pi} \frac{1}{b|\om|} \approx \frac{1}{\pi b} \frac{d \lambda}{\lambda}, \eenn
we find that the potential term renormalizes as
\benn
(c \Lambda(a^2-1)^2)_{\lambda} \Rightarrow (c \Lambda(a^2-1)^2)_{\lambda-d\lambda} \approx (c \Lambda)_{\lambda} \left[ (a_<^2 -1)^2 + (3 a_<^2 -1) \frac{2}{\pi b} \frac{d \lambda}{\lambda} \right], \label{ren} \eenn
\nid where we have ignored terms including higher powers of $d\lambda/\lambda$. To recast the new lower-frequency action into the same form as the original action the dimensionless coupling parameters must be renormalised. In particular, we observe that the dimensionless barrier height $c$ can either grow or shrink depending on the value of the dimensionless dissipation $b$. Note that the coordinates must also be rescaled (also known as wavefunction renormalization) for the potential in \eq{ren} to maintain the same quartic form as in \eq{dimpot}, thereby inducing a rescaling of $b$. We concentrate here on the renormalized potential coupling term and find that, up to a constant,
\benn
c_{\lambda-d\lambda}=c_{\lambda} \left[ 1 +\frac{d \lambda}{\lambda} \left(1-\frac{6}{\pi b} \right) \right], \eenn
\nid giving then the following differential RG flow equation
\benn
\frac{d c}{d \ln \lambda}=\left( \frac{b^*}{b} -1 \right) c. \lab{fl}
\eenn
\nid As the (dimensionless) barrier height $c$ renormalizes towards lower frequencies we observe two types of behavior depending on whether the parameter $b$ is greater or smaller than the critical value $b^*=6/\pi$ (the actual numerical value may well be slightly altered by going to higher orders in the perturbative expansion, but the important point to note that it is non-zero and thus gives rise to distinct dynamical phases). For $b > b^*$ the barrier height grows without bounds and thus effectively traps $\al(t)$ in one of the two minima, representing a localized phase. This localization can be brought about by increasing the magnitude of $\eta$, the numerical prefactor of our dissipative nonlocal priors, and/or increasing $\alb_1$ the distance between the two possible interpretations of the data. On the other hand, for $b < b^*$ the potential becomes ineffective in localizing $\al$, and thus $\al$ freely tunnels between the two wells, representing indeterminancy of the correct true model parameter.

It is interesting to note that a flow equation, similar to \eq{fl}, has been reported for the opposite limit (strong-coupling) using the instanton method[5,6]. Arguably what we have really shown is that even if one starts with weak coupling, so that it should be "easy" to jump from one interpretation to another, for $b > b^*$ we will flow to strong-coupling, at which point known results about localization take over.
  
\begin{figure}[h!]
\begin{center}
{\parbox{4cm}{\resizebox{6cm}{4cm}{\includegraphics{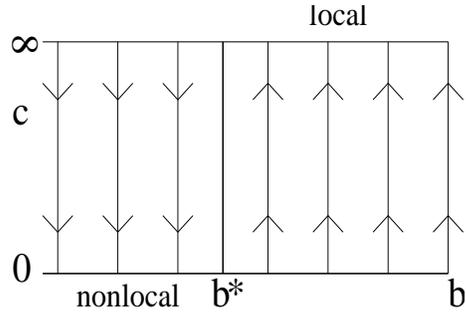}}}}
\end{center}
\caption{Schematic RG flow of the potential energy coupling parameter for $M \geq 2$. Note that the flow-lines are not expected to be strictly vertical due to wavefunction renormalization.} \lab{coup}
\end{figure}

The qualitative picture does not change when there are more than two possible model interpretations, $M>2$. In fact, the case of $M=\infty$ has been studied [7] where the potential energy landscape is taken to be sinusoidal, and it has been demonstrated that there again exists a critical value $b^*$ which separates a localized phase from a nonlocalized phase. The flow of the potential energy coupling constant $c$ is shown in Fig.\ref{coup} which is expected to be qualitatively correct across the whole range $2 \leq M \leq \infty$.

\section{Discussion}

In summary, the optimal model estimate in the response of ambiguous signals always results in random perceptual switching when the priors only constrain the local dynamics. We have shown that when we allow the possibility of $1/f$ noise in our priors then a specific model is learnt amongst the many possible models.

The connection between estimation theory and statistical mechanics is well known. One of the key results in statistical mechanics is that local interactions in one dimension can never lead to a phase transition. Thus if we are interested in, for example, learning a single parameter by making repeated observations, then there can be no phase transition to certainty about the value of this parameter as long as our prior hypotheses about its dynamics are equivalent to local models in statistical mechanics. Markov models, Gaussian processes with rational spectra, and other common priors all fall in this local class.

The common occurrence of $1/f$ fluctuations in nature motivates the analyses of estimation theory with such priors. Crucially, $1/f$ spectra do not correspond to local models. In fact they correspond exactly to the addition of friction to the path integral describing a quantum mechanical particle, a problem of general interest in condensed matter physics and more recently in quantum computing. Here we note one important consequence of these priors, namely that we can process data in a model which admits the possibility of time variation for the underlying parameter, but nonetheless find that our best estimate of this parameter is localized for all time to one of many equally plausible alternatives. It seems that $1/f$ priors may provide a way to understand the emergence of certainty more generally as a phase transition.

\subsubsection*{References}

\small{
[1] G. H. Fisher, Perception \& Psychophysics {\bf 4}, 189 (1968)

[2] E. de Boer, Handbk. Sens. Physiol. {\bf 3}, 479 (1976)

[3] W. Bialek and M. DeWeese, M. Phys. Rev. Lett. {\bf 74}, 3079 (1995)

[4] A. O. Caldeira and A. J. Leggett, Phys. Rev. Lett. {\bf 46}, 211 (1981)

[5] A. J. Bray and M. A. Moore, Phys. Rev. Lett {\bf 49}, 1545 (1982)

[6] A. J. Leggett, S. Chakravarty, A. T. Dorsey, M. P. A. Fisher, A. Garg and W. Zwerger, Rev. Mod. Phys. {\bf 59}, 1 (1987)

[7] M. P. A. Fisher and W. Zwerger, Phys. Rev. Lett {\bf 32}, 6190 (1985)
}

\end{document}